\documentclass[aps,twocolumn,a4paper,showpacs,nofootinbib,superscriptaddress]{revtex4}
\usepackage[dvips]{graphicx}

\begin{document}

\title{Electron-hole pairing in topological insulator thin film}
\author{D.K. Efimkin}
\affiliation{Institute for Spectroscopy, Russian Academy of Sciences, 142190 Troitsk, Moscow region, Russia}
\author{Yu.E. Lozovik}\email{lozovik@isan.troitsk.ru}
\affiliation{Institute for Spectroscopy, Russian Academy of Sciences, 142190 Troitsk, Moscow region, Russia}
\affiliation{Moscow Institute of Physics and Technology, 141700 Dolgoprudny, Moscow region, Russia}
\author{A.A. Sokolik}\email{aasokolik@yandex.ru}
\affiliation{Institute for Spectroscopy, Russian Academy of Sciences, 142190 Troitsk, Moscow region, Russia}
\begin{abstract}
We consider pairing of massless Dirac electrons and holes located on opposite surfaces of thin film of ``strong''
three-dimensional topological insulator. Such pairing was predicted to give rise to topological exciton condensate with
unusual properties. We estimate quantitatively achievable critical temperature of the pairing with taking into account
self-consistent screening of the Coulomb interaction, disorder and hybridization of electron and hole states caused by
a tunneling through the film. Increase of the gap above the hybridization value when the temperature is lowered can be
observable signature of the pairing. System parameters required to observe the electron-hole pairing are discussed.
\end{abstract}


\maketitle

\section{Introduction}

Research of nontrivial topological states of matter was highly stimulated in recent years by discovery of two- and
three-dimensional topological insulators (TIs) \cite{Hasan,Qi1}. Nonzero topological invariant characterizing global
topology of filled electron states in Hilbert space across the whole Brillouin zone distinguishes TI from a trivial
insulator. One of the most interesting properties of TI is the existence of gapless topologically protected edge (in a
two-dimensional case) or surface (in a three-dimensional case) electron states.

Unusual properties of surface modes manifest themselves most strikingly in second generation of ``strong''
three-dimensional TIs represented by such materials as $\mathrm{Bi}_2\mathrm{Se}_3$, $\mathrm{Bi}_2\mathrm{Te}_3$,
$\mathrm{Sb}_2\mathrm{Te}_3$, $\mathrm{Bi}_2\mathrm{Te}_2\mathrm{Se}$ and others \cite{ZhangLiu,Hsieh,Chen1,Xia}. In
these materials band structure of the surface states contains a Dirac cone, and electrons obey a two-dimensional
Dirac-Weyl equation for massless particles in the vicinity of a Dirac point. Electrons in graphene demonstrate similar
properties but have two inequivalent Dirac cones and additional two-fold degeneracy by spin projections
\cite{CastroNeto}.

Gap in a spectrum of the surface states can be induced when the time-reversal symmetry is broken by magnetic impurities
or in proximity to a ferromagnet \cite{Liu,Chen2,Qi2}. When the gap is opened in such a way on the whole surface of TI,
the spectacular topological magnetoelectric effect arises \cite{Qi2,Essin}. Another way to open the gap is to break the
gauge symmetry by a contact with superconductor, when the surface of TI acquires the properties of topological
superconductor \cite{FuKane}. Intrinsic Cooper pairing involving surface Dirac electrons induced by some mechanism can
also lead to analogue of topological superconductivity on a surface of TI \cite{Hor,Lu1,Diamantini}.

Such ``strong'' three-dimensional TIs as $\mathrm{Bi}_2\mathrm{Se}_3$, $\mathrm{Bi}_2\mathrm{Te}_3$ etc. have layered crystal structure
\cite{ZhangLiu,Chen1} with each layer consisting of five atomic layers (quintuple layer, QL) and having a thickness of about 1~nm. Fabrication of
thin films of these materials with arbitrary thickness, down to only one QL, was realized recently by means of epitaxial methods
\cite{G_Zhang,Li1,Li2,Bansal}, by vapor-solid growth \cite{Kong} and by mechanical exfoliation \cite{Hong,Shahil,Teweldebrhan}. Electronic properties
of TI thin films are strongly affected by a tunneling between opposite surfaces giving rise to a hybridization gap
\cite{Li2,Linder,Zhang,Sakamoto,Wang1,Park,Ebihara} and other observable phenomena, for example, unusual spin arrangement and splitting of the
zero-energy Landau level \cite{Lu,Zhang2,Yang,Zyuzin1}.

Chiral electrons on opposite surfaces of TI film constitute a strongly-coupled bilayer system which, in principle, can demonstrate various coherent
quantum phenomena. In the case of antisymmetric doping of these surfaces, one can realize a Coulomb-interaction mediated pairing of electrons on one
surface of the film and holes on the opposite surface \cite{Seradjeh1,Hao,Cho,Wang2,Mink1,MacDonald,Moon,Sodemann,Seradjeh2,Kim,Seradjeh3}, analogous
to that proposed earlier for massive electrons and holes in coupled semiconductor quantum wells
\cite{LozovikYudson,LozovikYudson2,LozovikBerman,Shevchenko}, and for massless Dirac electrons and holes in graphene bilayer
\cite{LozovikSokolik1,Min,Zhang1,Bistritzer,Kharitonov,LozovikSokolik2,Mink,LozovikOgarkov,Efimkin,LozovikRecent}. This kind of pairing is similar to
Cooper pairing of electrons in superconductors but occurs between spatially separated electrons and holes.

Although graphene, owing to its monoatomic thickness, allows to fabricate two-layer structures with record small
interlayer distance and demonstrates high carrier mobilities \cite{CastroNeto}, four-fold degeneracy of its electron
states leads to very strong screening of the pairing interaction and thus to rather low critical temperatures
\cite{Kharitonov,LozovikOgarkov}. In the case of TI films, predictions for critical temperature of the pairing can be
much more optimistic since electron states there have no degeneracy, as noted in \cite{Seradjeh1,Cho,Sodemann}.

Besides the weaker screening, the pairing in TI films could be more interesting from the viewpoint of the superfluid properties of resulting
topological ``exciton'' condensate. Zero-energy Majorana modes bound to vortex cores, gapless states on a contact with superconductor and other
interesting phenomena were predicted in such system \cite{Seradjeh1,Hao,Cho,Seradjeh2,Kim}. However whether the electron-hole pairing is achievable
in TI films in practice is not yet known. Critical temperature for the superfluid state about 100~K was estimated in \cite{Wang2,Kim}, however,
without taking into account the screening. In approximation of separable potential, the critical temperatures up to 0.1~K were obtained \cite{Mink1}.
The recent more complicated calculations including dynamical screening and correlation effects resulted in rather optimistic estimates (about
100-200~K) for the zero-temperature gap \cite{Sodemann}. For the quantum Hall regime, gaps of the order of hundreds of Kelvins were predicted
\cite{MacDonald}.

In our article, we study observable signatures of the electron-hole pairing in TI films in realistic conditions with
taking into account interaction screening, hybridization and disorder. We start with description of the pairing in
simple Bardeen-Cooper-Schrieffer (BCS) approximation (section~\ref{sec2}) and show that the critical temperature takes
practically observable values (at least 0.1~K) only at film thickness less than 15~nm. Hybridization between opposite
surfaces significantly affects the pairing at such small thicknesses, as studied in section~\ref{sec3}. Detection of
the pairing against the background of strong hybridization becomes the major issue here. We show that increase of the
gap due to the pairing as the temperature is lowered can be observed in TI films of moderate thickness of about 5-8~QL.

Our study of suppression of the pairing by disorder in section~\ref{sec4} shows the pairing requires rather high
carrier mobilities to be observable. In section~\ref{sec5} we demonstrate that the screening in a system with induced
gap is suppressed in comparison with the metallic-like screening assumed in the previous sections. By taking into
account this correlation effect we find out that predicted gap greatly increases in moderately thin films.

The calculations in this article refer to the case of $\mathrm{Bi}_2\mathrm{Se}_3$ films with $0.96\,\mbox{nm}$
thickness of one QL \cite{Bansal}. However our conclusions can be extended to other TIs of similar type with
controlling parameters (Fermi velocity, dielectric permittivity and hybridization gaps) close to those for
$\mathrm{Bi}_2\mathrm{Se}_3$.

\section{BCS approximation}\label{sec2}

\begin{figure}[t]
\begin{center}
\resizebox{0.7\columnwidth}{!}{\includegraphics{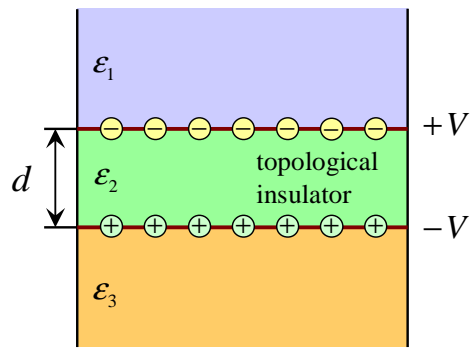}}
\end{center}
\caption{\label{Fig1}(Color online) TI film of thickness $d$ and dielectric permittivity $\varepsilon_2$ surrounded by
trivial insulators with permittivities $\varepsilon_1$ and $\varepsilon_3$. Electric potentials $\pm V$ on opposite
surfaces of the film create electron and hole gases of equal concentrations.}
\end{figure}

\begin{figure}[b]
\begin{center}
\resizebox{0.7\columnwidth}{!}{\includegraphics{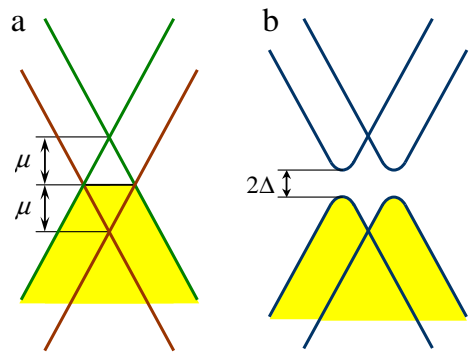}}
\end{center}
\caption{\label{Fig2}(Color online) Band picture of the pairing: (a) the applied potential difference shifts energies
of Dirac cones on opposite surfaces in opposite directions in such a way that electron and hole gases with chemical
potentials $\mu$ and $-\mu$ are formed; (b) the pairing opens the gap $2\Delta$ on the Fermi level.}
\end{figure}

One of the possible setups for the pairing is depicted in Fig.~\ref{Fig1}. The opposite electric potentials $\pm V$
imposed on the opposite surfaces of the TI film either by gate electrodes or by appropriate contacts cause
antisymmetric doping of these surfaces with electrons (top surfaces) and holes (bottom surface) up to the chemical
potentials $\mu$ and $-\mu$ respectively (Fig.~\ref{Fig2}). For simplicity, we assume here the electron-hole symmetry.
The alternative method to create electrons and holes can be based on chemical doping of the surfaces.

The inverted structure ``TI-insulator-TI'' was also proposed by several authors \cite{Hao,Cho,Wang2,Seradjeh2}, but we
shall demonstrate below that it is less suitable for the pairing.

Coulomb interaction between electron and hole residing on opposite surfaces of TI film undergoes combined screening by
three-dimensional dielectric environment and by two-dimensional electron and hole gases on these surfaces. We consider
the three-layer dielectric system (Fig.~1), where $\varepsilon_2$ is a dielectric permittivity of the TI film,
$\varepsilon_1$ and $\varepsilon_3$ are dielectric permittivities of insulators above and below the film. In such
system, the statically screened electron-hole interaction in the random phase approximation (RPA) is $-V(q)$, where
\begin{eqnarray}
V(q)=\frac{4\pi e^2}{qD(q)}\left[\varepsilon_2+\frac{4\pi e^2}q\Pi_{12}(q)\mathrm{sinh}\,qd\right]\label{V1}
\end{eqnarray}
(see also the similar formulas in \cite{LozovikYudson,MacDonald}). Here
\begin{eqnarray}
D(q)=(\varepsilon_1\varepsilon_3+\varepsilon_2^2)\,\mathrm{sinh}\,qd+
(\varepsilon_1+\varepsilon_3)\varepsilon_2\,\mathrm{cosh}\,qd\nonumber\\
-\frac{4\pi e^2}q\left[S_{11}(q)\Pi_{11}(q)+S_{22}(q)\Pi_{22}(q)+2\varepsilon_2\Pi_{12}(q)\right]\nonumber\\
+\frac{16\pi^2e^4}{q^2}\left[\Pi_{11}(q)\Pi_{22}(q)-\Pi_{12}^2(q)\right]\mathrm{sinh}\,qd,\label{V2}
\end{eqnarray}
\begin{eqnarray}
S_{11}(q)=\varepsilon_3\,\mathrm{sinh}\,qd+\varepsilon_2\,\mathrm{cosh}\,qd,\nonumber\\
S_{22}(q)=\varepsilon_1\,\mathrm{sinh}\,qd+\varepsilon_2\,\mathrm{cosh}\,qd,\label{V3}
\end{eqnarray}
$\Pi_{11}(q)$ and $\Pi_{22}(q)$ are static polarizabilities of electron and hole gases on top and bottom surfaces of
the film respectively, $\Pi_{12}(q)$ is the static anomalous interlayer polarizability; $d$ is the thickness of the
film.

We shall consider the pairing in the static approximation, i.e. with neglecting frequency dependencies of a gap and
pairing potential. At $0\leq q\leq 2p_\mathrm{F}$ the static polarizability of Dirac electron or hole gas is
$\Pi_0(q)=-g\mathcal{N}$ \cite{Wunsch,Hwang}, where $g$ is the degeneracy factor, $\mathcal{N}=\mu/2\pi v_\mathrm{F}^2$
is the density of states at the Fermi level, $p_\mathrm{F}=\mu/v_\mathrm{F}$ is the Fermi momentum, $v_\mathrm{F}$ is
the Fermi velocity of Dirac electrons and holes ($6.2\times10^5\,\mbox{m/s}$ for $\mathrm{Bi}_2\mathrm{Se}_3$
\cite{ZhangLiu}).

In this section we assume that the screening is the same as in intrinsic system without pairing, setting
$\Pi_{11}=\Pi_{22}=\Pi_0$, $\Pi_{12}=0$. The main advantage of TI film over graphene bilayer is the smaller degeneracy
factor $g=1$ (against $g=4$ for graphene) providing a weaker screening.

The static dielectric permittivity of strong three-dimensional TIs is rather large (e.g., $\varepsilon_2\approx80$ for
$\mathrm{Bi}_2\mathrm{Se}_3$ \cite{Bi2Se3}, $\varepsilon_2\approx30$ for $\mathrm{Bi}_2\mathrm{Te}_3$ \cite{Bi2Te3}),
which looks disappointing for realization of the pairing. However when the thickness of the film $d$ is much smaller
than the mean in-plane distance between electrons and holes (of the order of $p_\mathrm{F}^{-1}$), electric field lines
responsible for electron-hole interaction pass mainly through the media above and below the film. In this case the
screening is the same as in homogeneous medium with the dielectric permittivity $(\varepsilon_1+\varepsilon_3)/2$ and
does not depend on the dielectric permittivity $\varepsilon_2$ of the film itself (it follows directly from the limit
$qd\ll1$ of Eqs.~(\ref{V1})--(\ref{V3}) and was also noted in \cite{MacDonald}).

Generally, the pairing of massless Dirac fermions can be multi-band, when the both valence and conduction bands of both
layers are affected by the pairing correlations. The multi-band theory of the pairing in graphene bilayer provided
larger estimates for the gap and critical temperature than usual one-band BCS model both in static and dynamic
approximations \cite{LozovikSokolik2,Mink,LozovikOgarkov}. For the sake of simplicity, here we shall consider the
pairing in one-band BCS model, being aware, however, that our results for gap and critical temperature can be
underestimated.

The integral BCS equation for the gap function $\Delta(\mathbf{p})$ in the one-band approximation is the following
(analogously to that in \cite{LozovikSokolik1,LozovikSokolik2}):
\begin{eqnarray}
\Delta(\mathbf{p})=\int\frac{d\mathbf{p}'}{(2\pi)^2}\:\frac{1+\cos(\varphi-\varphi')}2\:V(|\mathbf{p}-\mathbf{p}'|)
\nonumber\\ \times\frac{\Delta(\mathbf{p'})}{2E(\mathbf{p}')}\tanh\frac{E(\mathbf{p}')}{2T},\label{gap_eq1}
\end{eqnarray}
where $\varphi$ and $\varphi'$ are azimuthal angles of the momenta $\mathbf{p}$ and $\mathbf{p}'$, entering the angular
factor, specific to chiral Dirac electrons, $E(\mathbf{p})=\sqrt{(v_\mathrm{F}p-\mu)^2+|\Delta(\mathbf{p})|^2}$ is the
Bogolyubov excitation energy, $T$ is the temperature.

According to the usual BCS recipe, we assume that the gap is nonzero and constant in some energy region of the
half-width $w$ around the Fermi surface:
\begin{eqnarray}
\Delta(\mathbf{p})=\left\{\begin{array}{rcl}\Delta,&\mbox{if}&|v_\mathbf{F}p-\mu|\leq w,\\
0,&\mbox{if}&|v_\mathbf{F}p-\mu|>w.\end{array}\right.\label{BCS}
\end{eqnarray}
Using (\ref{gap_eq1})--(\ref{BCS}) we can find the gap at $T=0$
\begin{eqnarray}
\Delta^\mathrm{BCS}_0=2w\,e^{-1/\lambda}\label{Delta0}
\end{eqnarray}
and the critical temperature
\begin{eqnarray}
T_\mathrm{c}=\frac{2we^\gamma}\pi\,e^{-1/\lambda}.\label{Tc}
\end{eqnarray}
Here $\gamma\approx0.577$ is the Euler constant, $\lambda$ is the dimensionless coupling constant calculated as an
average of interaction potential times the density of states over the Fermi surface:
\begin{eqnarray}
\lambda=\int\limits_0^{2\pi}\frac{d\varphi}{2\pi}\:\frac{1+\cos\varphi}2\:
\mathcal{N}V\left(2p_\mathrm{F}\sin\frac\varphi2\right).\label{lambda}
\end{eqnarray}

\begin{figure}[t]
\begin{center}
\resizebox{0.9\columnwidth}{!}{\includegraphics{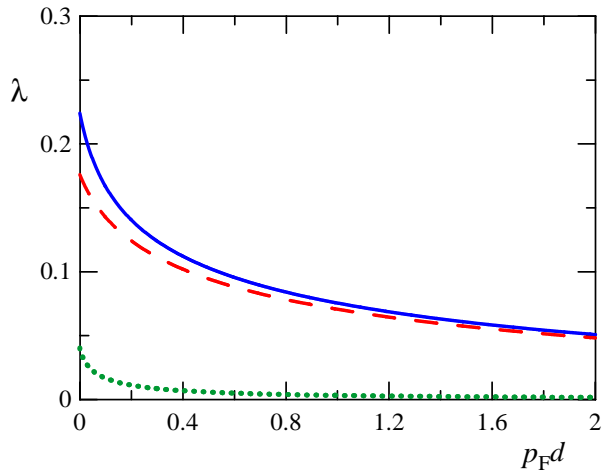}}
\end{center}
\caption{\label{Fig3}(Color online) The coupling constant (\ref{lambda}) for the pairing in
$\mathrm{Bi}_2\mathrm{Se}_3$ film with $\varepsilon_2=80$ as a function of $p_\mathrm{F}d$ at
$\varepsilon_1=\varepsilon_3=1$ (solid line), at $\varepsilon_1=\varepsilon_3=4$ (dashed line) and for TI-vacuum-TI
system with $\varepsilon_1=\varepsilon_3=80$, $\varepsilon_2=1$ (dotted line).}
\end{figure}

It is reasonable to take the pairing region half-width $w$ of the order of the chemical potential $\mu$, since there
are no other energy scales in the system (in contrast to superconductors, where $w$ can be taken of the order of Debye
frequency). Thus hereafter we take $w=\mu$. The maximal level of surface doping relative to the Dirac point in present
three-dimensional TIs, being limited by a position of the bulk valence band, is about $\mu\approx0.1\,\mbox{eV}$
\cite{ZhangLiu}. If we assume that the critical temperature (\ref{Tc}) should be at least $0.1\,\mathrm{K}$ for the
pairing to be observable, then at the maximal $w=\mu=0.1\,\mbox{eV}$ the coupling constant $\lambda$ should be not
smaller than $0.14$.

\begin{figure}[t]
\begin{center}
\resizebox{0.95\columnwidth}{!}{\includegraphics{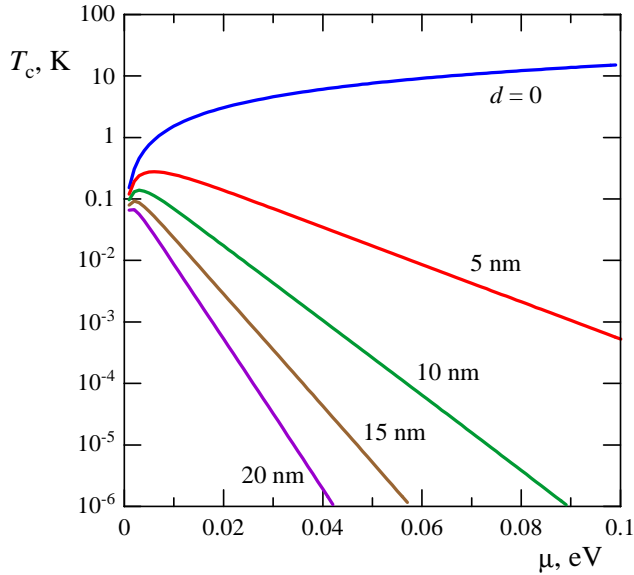}}
\end{center}
\caption{\label{Fig4}(Color online) The BCS critical temperature of the pairing (\ref{Tc}) in suspended
$\mathrm{Bi}_2\mathrm{Se}_3$ film as a function of the chemical potential $\mu$ at different film thicknesses $d$
indicated near the corresponding curves.}
\end{figure}

The coupling constants in the cases of suspended $\mathrm{Bi}_2\mathrm{Se}_3$ film with $\varepsilon_1=\varepsilon_3=1$
and of the film surrounded by a dielectric with $\varepsilon_1=\varepsilon_3=4$ are plotted in Fig.~\ref{Fig3} at
various $p_\mathrm{F}d$. It is seen that in the both cases $\lambda$ falls off very rapidly with increasing
$p_\mathrm{F}d$, and $\lambda>0.14$ requires $p_\mathrm{F}d<0.2$. We see also that strong screening by TI bilk
significantly suppresses the coupling constant in the inverted structure ``TI-insulator-TI'' even in the most favorable
case of the vacuum spacer with $\varepsilon_2=1$.

We can bring the coupling constant to its maximal value by approaching $\mu$ to zero and making the value of
$p_\mathrm{F}d$ arbitrarily small. But the preexponential factor in (\ref{Tc}), proportional to $\mu$, also decreases
in this case, thus we should not take too small $\mu$ in order to reach the highest $T_\mathrm{c}$. In Fig.~\ref{Fig4}
we plot $T_\mathrm{c}$ in a suspended $\mathrm{Bi}_2\mathrm{Se}_3$ film with $\varepsilon_1=\varepsilon_3=1$ as a
function of $\mu$ at different thicknesses $d$. It is seen that $T_\mathrm{c}$ is maximal at some nonzero $\mu$
dependent on $d$. In order to keep $T_\mathrm{c}$ above $0.1\,\mbox{K}$ we should take the film thickness $d$ not
exceeding $15\,\mbox{nm}$. The similar conclusion was made in \cite{Mink1}. However in this case a tunneling between
electron states on opposite surfaces of the film leads to significant hybridization of electron and hole states, which
will be considered in the next section.

\section{Influence of hybridization}\label{sec3}

Wave functions of gapless electron states on a surface of strong three-dimensional TI decay exponentially in the bulk
with characteristic depth of the order of several nanometers \cite{Lu,Ebihara}. In sufficiently thin films overlap of
wave functions of the states belonging to opposite surfaces of the film can occur. This results in avoided crossing of
dispersions of these states manifesting itself as opening of the hybridization gap $\Delta_\mathrm{T}$. This gap in the
spectrum is rather similar to the gap which could be opened by the pairing (see Fig.~\ref{Fig2}).

Indeed, the order parameter of the electron-hole pairing is $\langle a^{(1)}_\mathbf{p}b^{(2)}_{-\mathbf{p}}\rangle$,
where $a^{(1)}_\mathbf{p}$ is destruction operator for electron with momentum $\mathbf{p}$ on the top surface and
$b^{(2)}_{-\mathbf{p}}$ is destruction operator for hole with the opposite momentum $-\mathbf{p}$ on the bottom
surface. Electron-hole transformation implies that $b^{(2)}_{-\mathbf{p}}=a^{(2)+}_\mathbf{p}$, where
$a^{(2)+}_\mathbf{p}$ is creation operator for electron with momentum $\mathbf{p}$ on the bottom surface.

Therefore the pair of operators entering the order parameter $\langle
a^{(1)}_\mathbf{p}b^{(2)}_{-\mathbf{p}}\rangle$=$\langle a^{(1)}_\mathbf{p}a^{(2)+}_\mathbf{p}\rangle$ is analogous to
that describing the process of momentum-conserving electron tunneling between top and bottom surfaces. Thus the gap
equation for the total energy gap in the mean field approximation, describing both the pairing and hybridization,
differs from (\ref{gap_eq1}) by the additional term $\Delta_\mathrm{T}$ in the right-hand side:
\begin{eqnarray}
\Delta(\mathbf{p})=\Delta_\mathrm{T}+\int\frac{d\mathbf{p}'}{(2\pi)^2}\:\frac{1+\cos(\varphi-\varphi')}2\:
V(|\mathbf{p}-\mathbf{p}'|) \nonumber\\
\times\frac{\Delta(\mathbf{p'})}{2E(\mathbf{p}')}\tanh\frac{E(\mathbf{p}')}{2T}.\label{gap_eq2}
\end{eqnarray}
It is seen from (\ref{gap_eq2}) that the total gap $\Delta$ becomes larger than hybridization gap $\Delta_\mathrm{T}$
due to the pairing. We shall solve Eq.~(\ref{gap_eq2}), similarly to Eq.~(\ref{gap_eq1}), in the BCS-like approximation
(\ref{BCS}).

\begin{figure}[t]
\begin{center}
\resizebox{0.9\columnwidth}{!}{\includegraphics{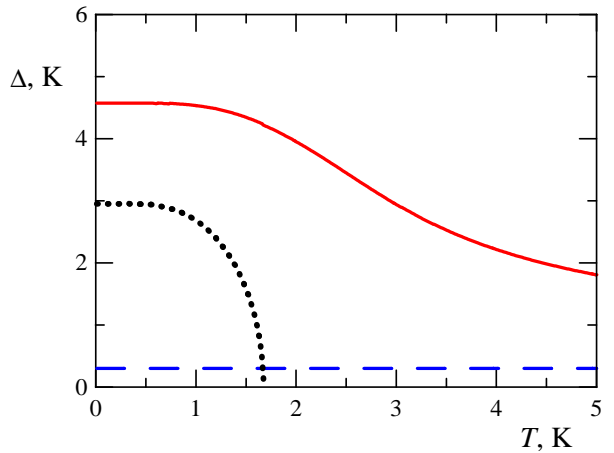}}
\end{center}
\caption{\label{Fig5}(Color online) Conceptual example of behavior of the gap $\Delta$ (solid line) as a function of
temperature $T$ in a system with hybridization at $\mu=0.1\,\mbox{eV}$, $\lambda=0.15$ and
$\Delta_\mathrm{T}=0.3\,\mbox{K}$. The initial hybridization gap $\Delta_\mathrm{T}$ (dashed line) and the BCS gap
calculated without hybridization (dotted line) are shown for comparison.}
\end{figure}

In Fig.~\ref{Fig5} the example of solution of Eq.~(\ref{gap_eq2}) is plotted under conditions when the hybridization
gap $\Delta_\mathrm{T}$ is approximately 10 times smaller than the zero-temperature BCS gap $\Delta^\mathrm{BCS}_0$
calculated without hybridization. In the absence of hybridization, the gap $\Delta$ would vanish at some critical
temperature, but nonzero hybridization makes $\Delta$ always nonzero and larger than $\Delta_\mathrm{T}$. At $T=0$ the
total gap is drastically increased even by weak hybridization --- due to nonlinearity of Eq.~(\ref{gap_eq2}) it is
larger than just a sum $\Delta_\mathrm{T}+\Delta^\mathrm{BCS}_0$; at $T\rightarrow\infty$ the gap gradually tends to
$\Delta_\mathrm{T}$. Thus the hybridization leads to the smearing of a phase transition into the paired state in close
analogy with behavior of a ferromagnet in external magnetic field.

How can we observe the pairing in the presence of hybridization? Besides possible superfluid signatures of the pairing
(see the Conclusions), we can still detect increase of the gap in the spectrum when the temperature is lowered
--- from the purely hybridization value $\Delta_\mathrm{T}$ at high temperatures to a somewhat larger value $\Delta_0$
at zero (or very low) temperature (Fig.~\ref{Fig5}). To be observable, this increase should be relatively large, i.e.
$\Delta_0-\Delta_\mathrm{T}$ should not be very small in comparison with $\Delta_\mathrm{T}$. In addition, the increase
of the gap should occur in reasonably narrow temperature range. This range can be estimated by a characteristic
temperature $T_\mathrm{char}$ at which the gap is halfway between $\Delta_\mathrm{T}$ and $\Delta_0$, i.e.
$\Delta(T_\mathrm{char})-\Delta_\mathrm{T}=[\Delta_0-\Delta_\mathrm{T}]/2$.

For calculations we take the data on hybridization gaps in $\mathrm{Bi}_2\mathrm{Se}_3$ films from the experiment
\cite{Zhang}, where the film thickness $d$ ranges from 2 to 5 QL, and from the theoretical paper \cite{Ebihara}, where
$d$ ranges from 1 to 16 QL (the close results for $\Delta_\mathrm{T}$ were also presented in the earlier paper
\cite{Linder}). It should be noted that in the overlapping region of $d$ from 2 to 5 QL the experimental gaps from
\cite{Zhang} are several times larger than the calculated gaps from \cite{Ebihara}, probably due to some impurities
enhancing the interlayer tunneling.

According to these data, we take several examples of $\mathrm{Bi}_2\mathrm{Se}_3$ films with different $d$ and
$\Delta_\mathrm{T}$. We take $\Delta_\mathrm{T}=0.126\,\mbox{eV}$ at $d=2\,\mbox{QL}$ and
$\Delta_\mathrm{T}=0.02\,\mbox{eV}$ at $d=5\,\mbox{QL}$ from \cite{Zhang} as examples of large hybridization gaps
(keeping in mind that the actual gap in the spectrum reported in the literature is $2\Delta_\mathrm{T}$). As examples
of small gaps, $\Delta_\mathrm{T}=6\times10^{-4}\,\mbox{eV}$ at $d=5\,\mbox{QL}$ and
$\Delta_\mathrm{T}=5\times10^{-5}\,\mbox{eV}$ at $d=8\,\mbox{QL}$ are taken from \cite{Ebihara}.

\begin{figure}[t]
\begin{center}
\resizebox{\columnwidth}{!}{\includegraphics{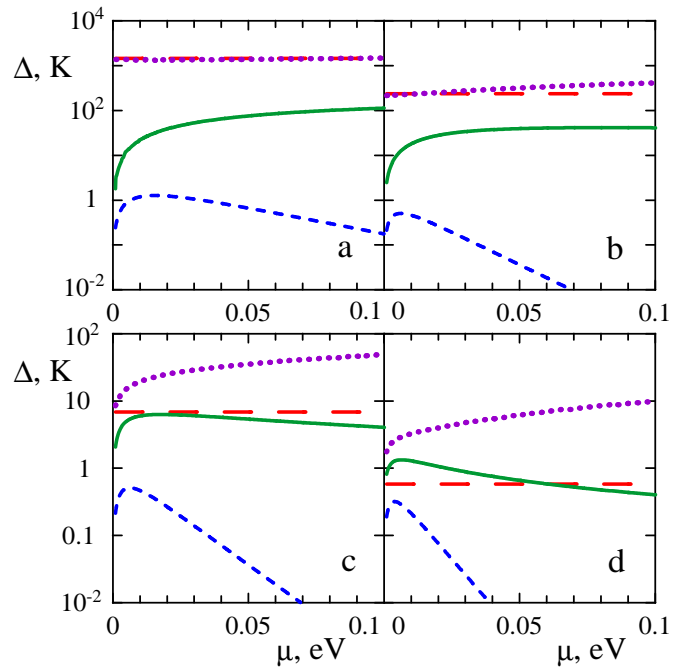}}
\end{center}
\caption{\label{Fig6}(Color online) Characteristics of the pairing in suspended $\mathrm{Bi}_2\mathrm{Se}_3$ films with
hybridization: increase $\Delta_0-\Delta_\mathrm{T}$ of the total gap (solid line) above the hybridization gap
$\Delta_\mathrm{T}$ (long-dashed line), characteristic temperature range $T_\mathrm{char}$ (dotted line) at which the
gap increases, and BCS gap $\Delta_0^\mathrm{BCS}$ (short-dashed line) calculated without hybridization. The data on
hybridization are taken from \cite{Zhang} for $d=2\,\mbox{QL}$ (a) and $5\,\mbox{QL}$ (b), and from \cite{Ebihara} for
$d=5\,\mbox{QL}$ (c) and $8\,\mbox{QL}$ (d).}
\end{figure}

The aforementioned characteristics of the pairing with hybridization are shown in Fig.~\ref{Fig6} for these four
examples of $\mathrm{Bi}_2\mathrm{Se}_3$ films at $\varepsilon_1=\varepsilon_3=1$. The gap calculated with
hybridization is several orders of magnitude larger than the BCS gap. In the cases of strong hybridization at small $d$
(Fig.~\ref{Fig6}(a,b)), the increase of the total gap above the purely hybridization gap $\Delta_\mathrm{T}$, being
negligible compared with $\Delta_\mathrm{T}$ itself and occurring in a range of several thousands Kelvins, can hardly
been observed. However when the hybridization is rather weak (Fig.~\ref{Fig6}(c,d)) the gap can grow significantly (up
to several times) when we lower the temperature by several tens of Kelvins. In this case the gap itself is not larger
than several Kelvins.

We can conclude that the temperature-dependent growth of the total gap is appreciable only if the hybridization is
sufficiently weak, i.e. in moderately thin films. Since the gap itself becomes too small in thicker films, the optimal
thickness for observing the pairing is about 5-8~QL.

\section{Influence of disorder}\label{sec4}

As known, charged impurities suppress an electron-hole pairing in two-layer system
\cite{LozovikYudson2,Bistritzer,Efimkin} since they, being usually uncorrelated in two layers, break the Cooper pairs
by differently scattering two pair constituents. According to the Abrikosov-Gor'kov theory, applied to the case of
electron-hole pairing, the Gor'kov equations remain the same as in the case of clean system, but with Matsubara
frequencies $\omega_n$ and gaps at these frequencies $\Delta_n$ renormalized in the following way:
\begin{eqnarray}
\omega_n\quad\rightarrow\quad\tilde\omega_n=\omega_n+(\gamma_{11}+\gamma_{22})\frac{\tilde\omega_n}{\sqrt{\tilde\omega_n^2+
\tilde\Delta_n^2}},\nonumber\\
\Delta_n\quad\rightarrow\quad\tilde\Delta_n=\Delta-2\gamma_{12}\frac{\tilde\Delta_n}{\sqrt{\tilde\omega_n^2+\tilde\Delta_n^2}},
\label{imp}
\end{eqnarray}
where $\omega_n=\pi T(2n+1)$ and $\Delta$ are initial Mathubara frequencies and gap. The quantities $\gamma_{ij}$
correspond to intralayer ($\gamma_{11}$ and $\gamma_{22}$ for top and bottom surfaces respectively) and interlayer
($\gamma_{12}$) correlation functions of random impurity potential.

The gap equation (\ref{gap_eq2}) for disordered system takes the form:
\begin{eqnarray}
\Delta=\Delta_\mathrm{T}+\lambda
T\sum_{n=-\infty}^{+\infty}\int\limits_{-w}^wd\xi\:\frac{\tilde\Delta_n}{\tilde\omega_n^2+\xi^2+\tilde\Delta_n^2}.
\label{gap_eq3}
\end{eqnarray}
When $\gamma_{11}+\gamma_{22}+2\gamma_{12}=0$, the impurities do not affect the result of energy integration and
frequency summation in (\ref{gap_eq3}) and thus the pairing characteristics remain the same. In superconductors it
corresponds to the case of nonmagnetic impurities which, according to the Anderson theorem, do not suppress the
pairing. In two-layer system this situation requires perfect anticorrelation of impurity potential between the layers
and thus is hardly realizable in practice.

In reality, when a range of impurity potential significantly exceeds a film thickness (the limit $d\rightarrow0$) and
both surfaces of the film are equally disordered, we have $\gamma_{11}=\gamma_{22}=\gamma_{12}=2\gamma$, where $\gamma$
is electron damping rate. For further calculations we consider the opposite case of short-range impurities and
relatively thick films ($d\rightarrow\infty$), when $\gamma_{11}=\gamma_{22}=2\gamma$, $\gamma_{12}=0$.

We can estimate $\gamma$ on the basis of the data on surface carrier mobilities $\mu_\mathrm{c}$ determining the
surface conductivity of TI, expressed through the Drude formula:
\begin{eqnarray}
\sigma=ne\mu_\mathrm{c}=\frac{e^2\mu}{4\pi\gamma},\label{sigma}
\end{eqnarray}
where $n=\mu^2/4\pi v_\mathrm{F}^2$ is the surface carrier concentration. In experiments on surface transport on
$\mathrm{Bi}_2\mathrm{Se}_3$ (see the references cited on pp.~1088-1089 of the review \cite{Qi1}) the measured
mobilities vary from $\sim100$ to $\sim20000\,\mbox{cm}^2/\mbox{V}\cdot\mbox{s}$. For calculations we choose the
example of dirty sample where $\mu_\mathrm{c}=500\,\mbox{cm}^2/\mbox{V}\cdot\mbox{s}$ and the example of clean sample
with $\mu_\mathrm{c}=10^4\,\mbox{cm}^2/\mbox{V}\cdot\mbox{s}$.

In usual Abrikosov-Gor'kov theory, the pair-breaking disorder reduces both the gap and critical temperature, and
sufficiently strong disorder can suppress the pairing completely. In our case the total gap $\Delta$ is always larger
than the hybridization gap $\Delta_\mathrm{T}$. However in the presence of disorder their difference
$\Delta-\Delta_\mathrm{T}$ diminishes in comparison with that in clean system.

\begin{figure}[t]
\begin{center}
\resizebox{\columnwidth}{!}{\includegraphics{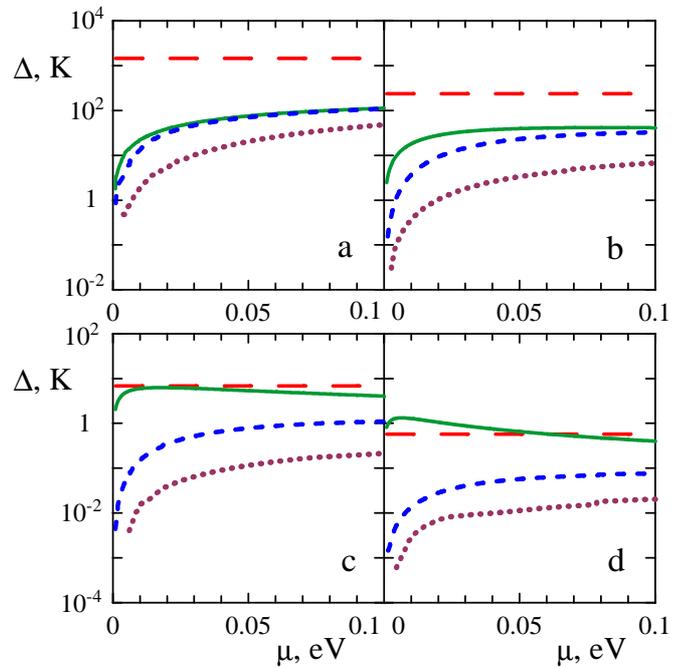}}
\end{center}
\caption{\label{Fig7}(Color online) Increase $\Delta_0-\Delta_\mathrm{T}$ of the zero temperature total gap above the
hybridization gap with no disorder (solid line), at $\mu_\mathrm{c}=10^4\,\mbox{cm}^2/\mbox{V}\cdot\mbox{s}$
(short-dashed line) and at $\mu_\mathrm{c}=500\,\mbox{cm}^2/\mbox{V}\cdot\mbox{s}$ (dotted line). The hybridization gap
(long-dashed line) in $\mathrm{Bi}_2\mathrm{Se}_3$ films is taken from \cite{Zhang} for $d=2\,\mbox{QL}$ (a) and
$5\,\mbox{QL}$ (b), and from \cite{Ebihara} for $d=5\,\mbox{QL}$ (c) and $8\,\mbox{QL}$ (d).}
\end{figure}

In Fig.~\ref{Fig7} the increase of the gap $\Delta_0-\Delta_\mathrm{T}$ calculated numerically from
(\ref{imp})--(\ref{gap_eq3}) is plotted at different disorder strengths for different examples of suspended
$\mathrm{Bi}_2\mathrm{Se}_3$ films. It is seen that disorder rather weakly affects the pairing at strong hybridization
(Fig.~\ref{Fig7}(a)) since the gap is very large in this case. At weak hybridization, when the effect of the pairing on
the gap is expected to be most pronounced, even moderate disorder drastically reduces $\Delta_0-\Delta_\mathrm{T}$
(Fig.~\ref{Fig7}(c,d)). Therefore we can conclude that the temperature-dependent increase of the gap due to the pairing
can be observed only in very clean (with carrier mobilities $\mu_\mathrm{c}>10^4\,\mbox{cm}^2/\mbox{V}\cdot\mbox{s}$)
and moderately thin (5-8~QL) TI films.

\section{Suppression of screening}\label{sec5}

In the previous sections we have considered the pairing potential screened by dielectric environment and by metallic
electron and hole gases on two surfaces of the film. However this consideration is not completely self-consistent since
the film becomes insulating on the surface due to appearance of the gap in the spectrum. When the gap is very large
(especially in the case of strong hybridization) the screening by electron and hole gases can substantially differ from
that in metallic system. In this case we should take into account self-consistent weakening of the screening caused by
the pairing (the similar effect was considered earlier for semiconductor quantum wells \cite{LozovikBerman} and
graphene bilayer \cite{Sodemann,LozovikRecent}).

When the gap $\Delta$ appears in the system due to interlayer pairing or hybridization, the intralayer static
polarizabilities $\Pi_{11}$ and $\Pi_{22}$ are no longer equal to the intrinsic polarizability $\Pi_0$ and can be
expressed in RPA as (see \cite{Sodemann,LozovikRecent})
\begin{eqnarray}
\Pi_{11,22}(q)=-g\sum_{\gamma\gamma'}\int\frac{d\mathbf{p}}{(2\pi)^2}\:
\frac{1+\gamma\gamma'\cos(\varphi-\varphi')}2\nonumber\\
\times\frac{u_{p\gamma}^2v_{p'\gamma'}^2+v_{p\gamma}^2u_{p'\gamma'}^2}{E_{p\gamma}+E_{p\gamma'}},\label{Pn}
\end{eqnarray}
where $\mathbf{p}'=\mathbf{p}+\mathbf{q}$, $\varphi$ and $\varphi'$ are azimuthal angles of the momenta $\mathbf{p}$
and $\mathbf{p}'$, $\gamma,\gamma'=\pm1$ are indices denoting the conduction ($+1$) and valence ($-1$) bands (i.e.
upper and lower parts of the double Dirac cone), $E_{p\gamma}=\sqrt{(\gamma v_\mathrm{F}p-\mu)^2+\Delta^2}$ is the
energy of Bogolyubov excitation in the band $\gamma$. The coherence factors $u_{p\gamma}$ and $v_{p\gamma}$ are
positive and determined by the equations:
\begin{eqnarray}
u^2_{p\gamma}=\frac12+\frac{\gamma v_\mathrm{F}p-\mu}{2E_{p\gamma}},\quad v^2_{p\gamma}=\frac12-\frac{\gamma
v_\mathrm{F}p-\mu}{2E_{p\gamma}}.
\end{eqnarray}

In a two-layer system, interlayer pairing or hybridization also leads to appearance of the anomalous polarizability
$\Pi_{12}$ describing direct response of charge density in one layer on electric field in the other layer. In RPA it
can be calculated as
\begin{eqnarray}
\Pi_{12}(q)=g\sum_{\gamma\gamma'}\int\frac{d\mathbf{p}}{(2\pi)^2}\:
\frac{1+\gamma\gamma'\cos(\varphi-\varphi')}2 \nonumber\\
\times\frac{2u_{p\gamma}v_{p\gamma}u_{p'\gamma'}v_{p'\gamma'}}{E_{p\gamma}+E_{p\gamma'}}.\label{Pa}
\end{eqnarray}

\begin{figure}[t]
\begin{center}
\resizebox{0.9\columnwidth}{!}{\includegraphics{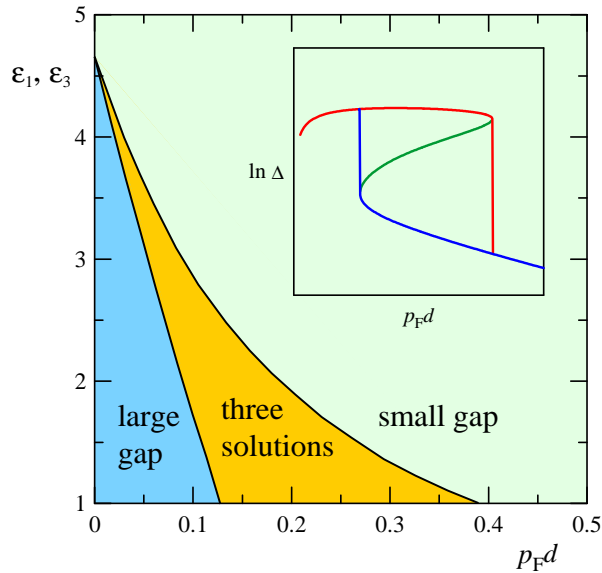}}
\end{center}
\caption{\label{Fig8}(Color online) Phase diagram of the pairing (without hybridization) in
$\mathrm{Bi}_2\mathrm{Se}_3$ film at different dielectric constants of surrounding medium $\varepsilon_1=\varepsilon_3$
and different dimensionless thicknesses $p_\mathrm{F}d$. The regions when the gap equation at $T=0$ gives one small
gap, three gaps and one large gap are shown. Inset: typical behavior of solutions $\Delta$ of the gap equation at
$\varepsilon_1,\varepsilon_3<4.5$.}
\end{figure}

\begin{figure}[t]
\begin{center}
\resizebox{\columnwidth}{!}{\includegraphics{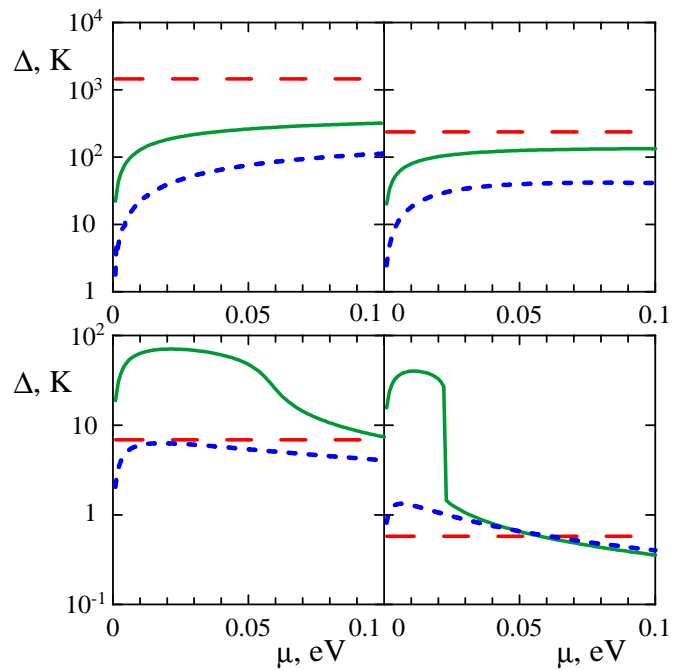}}
\end{center}
\caption{\label{Fig9}(Color online) Increase $\Delta_0-\Delta_\mathrm{T}$ of the zero temperature total gap in
suspended $\mathrm{Bi}_2\mathrm{Se}_3$ films above the hybridization gap with (solid line) and without (short-dashed
line) taking into account self-consistent suppression of the screening. The hybridization gap (long-dashed line) in
$\mathrm{Bi}_2\mathrm{Se}_3$ films is taken from \cite{Zhang} for $d=2\,\mbox{QL}$ (a) and $5\,\mbox{QL}$ (b), and from
\cite{Ebihara} for $d=5\,\mbox{QL}$ (c) and $8\,\mbox{QL}$ (d).}
\end{figure}

We performed numerical solving of the gap equation (\ref{gap_eq2}) with taking into account hybridization and with the pairing potential
(\ref{V1})--(\ref{V2}) self-consistently screened by electrons and holes with the polarizabilities (\ref{Pn}), (\ref{Pa}). In this case the usual BCS
method (\ref{BCS}) which reduces the pairing potential to a single coupling constant $\lambda$ is inapplicable since the integral in (\ref{lambda})
diverges due to absence of long-range screening in the gapped system. Thus we perform full integration of (\ref{gap_eq1}) over momentum $\mathbf{p}'$
in the region $|v_\mathrm{F}p'-\mu|<w$ where $\Delta(\mathbf{p}')\neq0$.

Our calculations show that at sufficiently weak coupling the gap is several times larger than in the case of metallic
screening. At stronger coupling two additional solutions of the gap equation appear, which are by several orders of
magnitude larger than the small gap that existed at weak coupling. Only the maximal of the resulting three gaps,
providing the lowest ground state energy, will be established in the system. At further increase of the coupling (say,
decrease of $p_\mathrm{F}d$) two smallest solutions disappear and only one large gap remains (see the inset in
Fig.~\ref{Fig8}).

In Fig.~\ref{Fig8} we show the phase diagram of $\mathrm{Bi}_2\mathrm{Se}_3$ film at $T=0$ with self-consistent
weakening of the screening but without hybridization. At nonzero hybridization the regions of one large gap and of
three solutions in this diagram would grow in size. At nonzero temperature the smallest gap gradually vanishes at small
critical temperature (as in usual BCS model), but two largest gaps disappear at much larger temperature abruptly from
nonzero values to zero, which is characteristic of the first-order phase transition (this fact was also noted in
\cite{Sodemann}).

In Fig.~\ref{Fig9} we demonstrate that $\Delta_0-\Delta_\mathrm{T}$ becomes larger in several times at strong
hybridization (Fig.~\ref{Fig9}(a,b)) and by the order of magnitude at weaker hybridization (Fig.~\ref{Fig9}(c,d)) in
comparison with the results obtained with metallic screening. Our predictions for observability of the pairing become
more optimistic, when we treat the screening self-consistently, especially for relatively thick films with weak
hybridization, where even sharp transition to a strongly-correlated state can occur (Fig.~\ref{Fig9}(d)).

\section{Conclusions}

We have considered the pairing of spatially separated massless Dirac electrons and holes created on opposite surfaces
of TI thin film by antisymmertic doping. The main advantage of such system over two-layer graphene system is four times
smaller degeneracy of electron states leading to much weaker screening and much larger coupling constants. The effect
of large bulk dielectric constant of the TI itself on the pairing can be negligible when the film is sufficiently thin.

Our calculations in BCS approximation show that the pairing gap is large enough to be observable when the film
thickness is less than $15\,\mbox{nm}$. In this case, however, tunneling between opposite surfaces leads to
hybridization of electron and hole states. Such hybridization, on the one hand, effectively increases the gap. On the
other hand, the pairing can become hardly observable on background of strong hybridization.

We show that the pairing causes increase of the total gap in the spectrum above the purely hybridization gap when the
temperature is decreased. This effect can be observed only in moderately thin films (about 5-8~QL for
$\mathrm{Bi}_2\mathrm{Se}_3$) where the hybridization is rather weak and effect of the pairing is noticeable on its
background. Another way to observe the pairing can be based on measuring of change of the gap with varying surface
electron and hole chemical potentials. Charged impurities and other disorder on surfaces of the film suppress the
pairing, so as it can be observed only in sufficiently clean TIs (surface carrier mobility should be at least
$10^4\,\mbox{cm}^2/\mbox{V}\cdot\mbox{s}$ by the order of magnitude).

Also we have demonstrated that BCS approximation, by assuming the metallic screening by surface carriers,
underestimates the coupling strength since the real screening in the gapped system is much weaker. The self-consistent
treatment of the screening demonstrates that the gap can be by orders of magnitude larger than given by BCS model, and
observation of the pairing in TI films can turn out to be more feasible. Multi-band and dynamical effects can
additionally increase the gap.

Realization of electron-hole pairing in TI thin film is a difficult task including chemical doping of TI bulk to
insulating state and, at the same time, doping of two opposite surfaces to electron- and hole-metallic states by means
of gate electrodes or charged impurities. Another challenge is fabrication of separate contacts to the surfaces.
However considerable progress achieved recently in experimental studies of TIs and TI thin films (see \cite{Hasan,Qi1}
and references therein) gives hope that conditions suitable for the electron-hole pairing can be reached in nearest
years. In particular, transport measurements in a regime of coexistence of electrons and holes on opposite surfaces of
$\mathrm{Bi}_2\textrm{Se}_3$ film were carried out in Ref.~\cite{Chen}.

In this article, we considered only the manifestation of the pairing in the temperature dependence of the gap in the
spectrum. For exciton condensate in electron-hole bilayers, such signatures as dipolar superfluidity
\cite{LozovikYudson}, Josephson-like effects \cite{LozovikYudson,Shevchenko,Klyuchnik,Poushnov}, peculiarities of the
drag effect \cite{Conti,Hu,Mink1} and anomalous electromagnetic response \cite{Ovchinnikov,Balatsky1} were predicted.
Strong hybridization occurring in thin TI films imposes serious limitations on observability of these phenomena. As
known, the tunneling in electron-hole bilayer leads to fixation of the condensate phase and to absence of uniform
dipolar superfluidity. However the superfluidity can arise locally in a form of Josephson-like vortices
\cite{Klyuchnik,Poushnov} or vortex lattice \cite{Shevchenko}. In the latter case the dipolar current flowing along the
bilayer should exceed some critical value in order for the vortex lattice to be stable. At high enough temperature the
vortex lattice undergoes a dislocation-mediated melting when pairs of edge dislocations dissociate and proliferate (see
\cite{Shevchenko}). It is interesting to note that dislocation ends in this case are topologically equivalent to
vortices in a superfluid condensate and thus can settle Majorana zero energy modes \cite{Seradjeh1} which can be
manipulated for the purpose of quantum computations.

The work was supported by Russian Foundation for Basic Research. D.K.E. and A.A.S. were also supported by the Dynasty Foundation and by the grant of
the President of Russian Federation MK-5288.2011.2.


\begin{thebibliography}{99}

\bibitem{Hasan}
M.Z. Hasan, C.L. Kane, Rev. Mod. Phys. \textbf{82}, 3045 (2010).

\bibitem{Qi1}
X.-L. Qi, S.-C. Zhang, Rev. Mod. Phys. \textbf{83}, 1057 (2011).

\bibitem{ZhangLiu}
H. Zhang, C.-X. Liu, X.-L. Qi, X. Dai, Z. Fang, S.-C. Zhang, Nature Phys. \textbf{5}, 438 (2009).

\bibitem{Hsieh}
D. Hsieh, Y. Xia, L. Wray, D. Qian, A. Pal, J.H. Dil, J. Osterwalder, F. Meier, G. Bihlmayer, C.L. Kane, Y.S. Hor, R.J.
Cava, M.Z. Hasan, Science \textbf{323}, 919 (2009).

\bibitem{Chen1}
Y.L. Chen, J.G. Analytis, Z.H. Chu, Z.K. Liu, S.K. Mo, X.L. Qi, H.J. Zhang, D.H. Lu, X. Dai, Z. Fang, S.C. Zhang, I.R.
Fisher, Z. Hussain, Z.X. Shen, Science \textbf{325}, 178 (2009).

\bibitem{Xia}
Y. Xia, L. Wray, D. Qian, D. Hsieh, A. Pal, H. Lin, A. Bansil, D. Grauer, Y. Hor, R. Cava, M.Z. Hasan, Nature Phys.
\textbf{5}, 398 (2009).

\bibitem{CastroNeto}
A.H. Castro Neto, F. Guinea, N.M.R. Peres, K.S. Novoselov, A.K. Geim, Rev. Mod. Phys. \textbf{81}, 109 (2009).

\bibitem{Liu}
Q. Liu, C.-X. Liu, C. Xu, X.-L. Qi, S.-C. Zhang, Phys. Rev. Lett. \textbf{102}, 156603 (2009).

\bibitem{Chen2}
Y.L. Chen, J.-H. Chu, J.G. Analytis, Z.K. Liu, K. Igarashi, H.-H. Kuo, X.L. Qi, S.K. Mo, R.G. Moore, D.H. Lu, M.
Hashimoto, T. Sasagawa, S.C. Zhang, I.R. Fisher, Z. Hussain, Z.X. Shen, Science \textbf{329}, 659 (2010).

\bibitem{Qi2}
X.L. Qi, T.L. Hughes, S.C. Zhang, Phys. Rev. B \textbf{78}, 195424 (2008).

\bibitem{Essin}
A.M. Essin, J.E. Moore, D. Vanderbilt, Phys. Rev. Lett. \textbf{102}, 146805 (2009).

\bibitem{FuKane}
L. Fu, C.L. Kane, Phys. Rev. Lett. \textbf{100}, 096407 (2008).

\bibitem{Hor}
Y.S. Hor, A.J. Williams, J.G. Checkelsky, P. Roushan, J. Seo, Q. Xu, H.W. Zandbergen, A. Yazdani, N.P. Ong, R.J. Cava,
Phys. Rev. Lett. \textbf{104}, 057001 (2010).

\bibitem{Lu1}
C.-K. Lu, I.F. Herbut, Phys. Rev. B \textbf{82}, 144505 (2010).

\bibitem{Diamantini}
M.C. Diamantini, P. Sodano, C.A. Trugenberger, New J. Phys. \textbf{14}, 063013 (2012).

\bibitem{G_Zhang}
G. Zhang, H. Qin, J. Teng, J. Guo, Q. Guo, X. Dai, Z. Fang, K. Wu, Appl. Phys. Lett. \textbf{95}, 053114 (2009).

\bibitem{Li1}
H.D. Li, Z.Y. Wang, X. Kan, X. Guo, H.T. He, Z. Wang, J.N. Wang, T.L. Wong, N. Wang, M.H. Xie, New J. Phys.
\textbf{12}, 103038 (2010).

\bibitem{Li2}
Y.-Y. Li, G. Wang, X.-G. Zhu, M.-H. Liu, C. Ye, X. Chen, Y.-Y. Wang, K. He, L.-L. Wang, X.-C. Ma, H.-J. Zhang, X. Dai,
Z. Fang, X.-C. Xie, Y. Liu, X.-L. Qi, J.-F. Jia, S.-C. Zhang, Q.-K. Xue, Adv. Mater. \textbf{22}, 4002 (2010).

\bibitem{Bansal}
N. Bansal, Y.S. Kim, E. Edrey, M. Brahlek, Y. Horibe, K. Iida, M. Tanimura, G.-H. Li, T. Feng, H.-D. Lee, T.
Gustafsson, E. Andrei, S. Oh, Thin Solid Films \textbf{520}, 224 (2010).

\bibitem{Kong}
D. Kong, W. Dang, J.J. Cha, H. Li, S. Meister, H. Peng, Z. Liu, Y. Cui, Nano Lett. \textbf{10}, 2245 (2010).

\bibitem{Hong}
S.S. Hong, W. Kundhikanjana, J.J. Cha, K. Lai, D. Kong, S. Meister, M.A. Kelly, Z.-X. Shen, Y. Cui, Nano Lett.
\textbf{10}, 3118 (2010).

\bibitem{Shahil}
K.M.F. Shahil, M.Z. Hossain, D. Teweldebrhan, A.A. Balandin, Appl. Phys. Lett. \textbf{96}, 153103 (2010).

\bibitem{Teweldebrhan}
D. Teweldebrhan, V. Goyal, A.A. Balandin, Nano Lett. \textbf{10}, 1209 (2010).

\bibitem{Linder}
J. Linder, T. Yokoyama, A. Sudbo, Phys. Rev. B \textbf{80}, 205401 (2009).

\bibitem{Zhang}
Y. Zhang, K. He, C.-Z. Chang, C.-L. Song, L.-L. Wang, X. Chen, J.-F. Jia, Z. Fang, X. Dai, W.-Y. Shan, S.-Q. Shen, Q.
Niu, X.-L. Qi, S.-C. Zhang, X.-C. Ma, Q.-K. Xue, Nature Phys. \textbf{6}, 584 (2010).

\bibitem{Sakamoto}
Y. Sakamoto, T. Hirahara, H. Miyazaki, S.I. Kimura, S. Hasegawa, Phys. Rev. B \textbf{81}, 165432 (2010).

\bibitem{Wang1}
G. Wang, X. Zhu, J. Wen, X. Chen, K. He, L. Wang, X. Ma, Y. Liu, X. Dai, Z. Fang, J. Jia, Q. Xue, Nano Research
\textbf{3}, 874 (2010).

\bibitem{Park}
K. Park, J.J. Heremans, V.W. Scarola, D. Minic, Phys. Rev. Lett. \textbf{105}, 186801 (2010).

\bibitem{Ebihara}
K. Ebihara, K. Yada, A. Yamakage, Y. Tanaka, Physica E \textbf{44}, 885 (2012).

\bibitem{Lu}
H.-Z. Lu, W.-Y. Shan, W. Yao, Q. Niu, S.-Q. Shen, Phys. Rev. B \textbf{81}, 115407 (2010).

\bibitem{Zhang2}
X. Zhang, J. Wang, S.-C. Zhang, Phys. Rev. B \textbf{82}, 245107 (2010).

\bibitem{Yang}
Z. Yang, J.H. Han, Phys. Rev. B \textbf{83}, 045415 (2011).

\bibitem{Zyuzin1}
A.A. Zyuzin, A.A. Burkov, Phys. Rev. B \textbf{83}, 195413 (2011).

\bibitem{Seradjeh1}
B. Seradjeh, J.E. Moore, M. Franz, Phys. Rev. Lett. \textbf{103}, 066402 (2009).

\bibitem{Hao}
N. Hao, P. Zhang, Y. Wang, Phys. Rev. B \textbf{84}, 155447 (2011).

\bibitem{Cho}
G.Y. Cho, J.E. Moore, Phys. Rev. B \textbf{84}, 165101 (2011).

\bibitem{MacDonald}
D. Tilahun, B. Lee, E.M. Hankiewicz, A.H. MacDonald, Phys. Rev. Lett. \textbf{107}, 246401 (2011).

\bibitem{Wang2}
Z. Wang, N. Hao, Z.-G. Fu, P. Zhang, New J. Phys. \textbf{14}, 063010 (2012).

\bibitem{Mink1}
M.P. Mink, H.T.C. Stoof, R.A. Duine, M. Polini, G. Vignale, Phys. Rev. Lett. \textbf{108}, 186402 (2012).

\bibitem{Moon}
E.G. Moon, C. Xu, Europhys. Lett. \textbf{97}, 66008 (2012).

\bibitem{Sodemann}
I. Sodemann, D.A. Pesin, A.H. MacDonald, Phys. Rev. B \textbf{85}, 195136 (2012).

\bibitem{Seradjeh2}
B. Seradjeh, ArXiv:cond-mat/1203.6628.

\bibitem{Kim}
Y. Kim, E.M. Hankiewicz, M.J. Gilbert, ArXiv:cond-mat/1204.6351.

\bibitem{Seradjeh3}
B. Seradjeh, Phys. Rev. B \textbf{85}, 235146 (2012).

\bibitem{LozovikYudson}
Yu.E. Lozovik, V.I. Yudson, Pisma v ZhETF \textbf{22}, 556 (1975) [JETP Lett. \textbf{22}, 274 (1975)]; Zh. Eksp. Teor.
Fiz. \textbf{71}, 738 (1976) [Sov. Phys. JETP \textbf{44}, 389 (1976)].

\bibitem{LozovikYudson2}
Yu.E. Lozovik, V.I. Yudson, Solid State Commun. \textbf{21}, 211 (1977).

\bibitem{LozovikBerman}
Yu.E. Lozovik, O.L. Berman, Zh. Eksp. Teor. Fiz. \textbf{111}, 1879 (1997) [JETP \textbf{84}, 1027 (1997)].

\bibitem{Shevchenko}
S.I. Shevchenko, Phys. Rev. Lett. \textbf{72}, 3242 (1994).

\bibitem{LozovikSokolik1}
Yu.E. Lozovik, A.A. Sokolik, Pis'ma v ZhETF \textbf{87}, 61 (2008) [JETP Lett. \textbf{87}, 55 (2008)].

\bibitem{Min}
H. Min, R. Bistritzer, J.-J. Su, A.H. MacDonald, Phys. Rev. B \textbf{78}, 121401(R) (2008).

\bibitem{Zhang1}
C.-H. Zhang, Y.N. Joglekar, Phys. Rev. B \textbf{77}, 233405 (2008).

\bibitem{Bistritzer}
R. Bistritzer, A.H. MacDonald, Phys. Rev. Lett. \textbf{101}, 256406 (2008).

\bibitem{Kharitonov}
M.Yu. Kharitonov, K.B. Efetov, Phys. Rev. B \textbf{78}, 241401(R) (2008);
Semicond. Sci. Tech. \textbf{25}, 034004 (2010).

\bibitem{LozovikSokolik2}
Yu.E. Lozovik, A.A. Sokolik, Phys. Lett. A \textbf{374}, 326 (2009); Eur. Phys. J. B \textbf{73}, 195 (2009).

\bibitem{LozovikOgarkov}
Yu.E. Lozovik, S.L. Ogarkov, A.A. Sokolik, Philos. Trans. Roy. Soc. A \textbf{368}, 5417 (2010).

\bibitem{Mink}
M.P. Mink, H.T.C. Stoof, R.A. Duine, A.H. MacDonald, Phys. Rev. B \textbf{84}, 155409 (2011).

\bibitem{Efimkin}
D.K. Efimkin, V.A. Kulbachinskii, Yu.E. Lozovik, Pis'ma v ZhETF \textbf{93}, 238 (2011) [JETP Lett. \textbf{93}, 219
(2011)].

\bibitem{LozovikRecent}
Yu.E. Lozovik, S.L. Ogarkov, A.A. Sokolik, ArXiv:cond-mat/1202.4978v2.

\bibitem{Wunsch}
B. Wunsch, T. Stauber, F. Sols, F. Guinea, New J. Phys. \textbf{8}, 318 (2006).

\bibitem{Hwang}
E.H. Hwang, S. Das Sarma, Phys. Rev. B \textbf{75}, 205418 (2007).

\bibitem{Bi2Se3}
Collaboration: Authors and editors of the volumes III/17E-17F-41C: Bismuth selenide (Bi2Se3) optical properties,
dielectric constants. Madelung, O., Rossler, U., Schulz, M. (ed.). SpringerMaterials --- The Landolt-Bornstein Database
(http://www.springermaterials.com). \verb"DOI: 10.1007/10681727_945"

\bibitem{Bi2Te3}
Collaboration: Authors and editors of the volumes III/17E-17F-41C: Bismuth telluride (Bi2Te3) optical properties,
dielectric constant. Madelung, O., Rossler, U., Schulz, M. (ed.). SpringerMaterials --- The Landolt-Bornstein Database
(http://www.springermaterials.com). \verb"DOI: 10.1007/10681727_963"

\bibitem{Klyuchnik}
A.V. Klyuchnik, Yu.E. Lozovik, Zh. Eksp. Teor. Fiz. \textbf{76}, 670 (1978) [Sov. Phys. JETP \textbf{49}, 335 (1978)].

\bibitem{Chen}
J. Chen, X.Y. He, K.H. Wu, Z.Q. Ji, L. Lu, J.R. Shi, J.H. Smet, Y.Q. Li, Phys. Rev. B \textbf{83}, 241304(R) (2011).

\bibitem{Poushnov}
Yu.E. Lozovik, A.V. Poushnov, Phys. Lett. A \textbf{228}, 399 (1997).

\bibitem{Conti}
S. Conti, G. Vignale, A.H. MacDonald, Phys. Rev. B \textbf{57}, R6846 (1998).

\bibitem{Hu}
B.Y.-K. Hu, Phys. Rev. Lett. \textbf{85}, 820 (2000).

\bibitem{Ovchinnikov}
Yu.E. Lozovik, I.V. Ovchinnikov, Phys. Rev. B \textbf{66}, 075124 (2002).

\bibitem{Balatsky1}
A.V. Balatsky, Y.N. Joglekar, P.B. Littlewood, Phys. Rev. Lett. \textbf{93}, 266801 (2004).

\end{thebibliography}
\end{document}